\documentclass[a4paper,11pt]{article}
\pdfoutput=1 

\usepackage{jcappub} 

\usepackage[T1]{fontenc} 
\usepackage{hyperref}%
\usepackage{orcidlink}

\def\ref@jnl#1{{\jnl@style#1}}%
\makeatletter
\providecommand{\ref@jnl}[1]{#1}
\newcommand\aj{\ref@jnl{AJ}}
\newcommand\psj{\ref@jnl{PSJ}}
\newcommand\araa{\ref@jnl{ARA\&A}}
\providecommand\apj{\ref@jnl{ApJ}}
\newcommand\apjl{\ref@jnl{ApJL}}     
\newcommand\apjs{\ref@jnl{ApJS}}
\newcommand\apss{\ref@jnl{Ap\&SS}}
\newcommand\aap{\ref@jnl{A\&A}}
\newcommand\aapr{\ref@jnl{A\&A~Rv}}
\newcommand\aaps{\ref@jnl{A\&AS}}
\newcommand\azh{\ref@jnl{AZh}}
\newcommand\baas{\ref@jnl{BAAS}}
\newcommand\icarus{\ref@jnl{Icarus}}
\newcommand\jaavso{\ref@jnl{JAAVSO}}  
\newcommand\jrasc{\ref@jnl{JRASC}}
\newcommand\memras{\ref@jnl{MmRAS}}
\newcommand\mnras{\ref@jnl{MNRAS}}
\providecommand\pra{\ref@jnl{PhRvA}}
\providecommand\prb{\ref@jnl{PhRvB}}
\providecommand\prc{\ref@jnl{PhRvC}}
\providecommand\prd{\ref@jnl{PhRvD}}
\providecommand\pre{\ref@jnl{PhRvE}}
\providecommand\prl{\ref@jnl{PhRvL}}
\newcommand\pasp{\ref@jnl{PASP}}
\newcommand\pasj{\ref@jnl{PASJ}}
\newcommand\qjras{\ref@jnl{QJRAS}}
\newcommand\skytel{\ref@jnl{S\&T}}
\newcommand\solphys{\ref@jnl{SoPh}}
\newcommand\sovast{\ref@jnl{Soviet~Ast.}}
\newcommand\ssr{\ref@jnl{SSRv}}
\newcommand\zap{\ref@jnl{ZA}}
\newcommand\iaucirc{\ref@jnl{IAUC}}
\newcommand\aplett{\ref@jnl{Astrophys.~Lett.}}
\newcommand\apspr{\ref@jnl{Astrophys.~Space~Phys.~Res.}}
\newcommand\bain{\ref@jnl{BAN}}
\newcommand\fcp{\ref@jnl{FCPh}}
\newcommand\gca{\ref@jnl{GeoCoA}}
\newcommand\grl{\ref@jnl{Geophys.~Res.~Lett.}}
\newcommand\jgr{\ref@jnl{J.~Geophys.~Res.}}
\newcommand\jqsrt{\ref@jnl{JQSRT}}
\newcommand\memsai{\ref@jnl{MmSAI}}
\newcommand\nphysa{\ref@jnl{NuPhA}}
\newcommand\physrep{\ref@jnl{PhR}}
\newcommand\physscr{\ref@jnl{PhyS}}
\newcommand\planss{\ref@jnl{Planet.~Space~Sci.}}
\newcommand\procspie{\ref@jnl{Proc.~SPIE}}
\newcommand\actaa{\ref@jnl{AcA}}
\newcommand\caa{\ref@jnl{ChA\&A}}
\newcommand\cjaa{\ref@jnl{ChJA\&A}}
\newcommand\jcap{\ref@jnl{JCAP}}
\newcommand\na{\ref@jnl{NewA}}
\newcommand\nar{\ref@jnl{NewAR}}
\newcommand\pasa{\ref@jnl{PASA}}
\newcommand\rmxaa{\ref@jnl{RMxAA}}
\newcommand\maps{\ref@jnl{M\&PS}}
\newcommand\aas{\ref@jnl{AAS Meeting Abstracts}}
\newcommand\dps{\ref@jnl{AAS/DPS Meeting Abstracts}}
\makeatother

\usepackage{placeins}
\usepackage{xspace}
\usepackage{acronym}

\acrodef{SNR}{signal-to-noise ratio}
\newcommand{\snr}{\ac{SNR}\xspace}
\newcommand{\lgnrm}{\texttt{lognormal\allowbreak\_galaxies}\xspace}\newcommand{\mice}{{\small \textsc{micecat}}\xspace}
\newcommand{\kmsMpc}{\ensuremath{\mbox{km s}^{-1} \,\mbox{Mpc}^{-1}}\xspace}

\newcommand{\CU}{Department of Physics, Cotton University, Panbazar, Guwahati 781001, Assam, India}
\newcommand{\ICRR}{KAGRA Observatory, Institute for Cosmic Ray Research, The University of Tokyo, 5-1-5, Kashiwanoha, Kashiwa, Chiba 277-8583, Japan}
\newcommand{\IUCAA}{Inter-University Centre for Astronomy and Astrophysics, Post Bag 4, Ganeshkhind, Pune 411 007, India}
\newcommand{\IPMU}{Kavli Institute for the Physics and Mathematics of the Universe (WPI), The University of Tokyo, 5-1-5 Kashiwanoha, Kashiwa, Chiba 277-8583, Japan}

\title{\boldmath Comparing the Detectability of the Baryon Acoustic Oscillation through Gravitational-Wave Autocorrelation and Cross-Correlation with Galaxies}

\author[a]{Gautam Bhuyan,\orcidlink{0000-0003-2428-037X}}
\author[b]{Tathagata Ghosh,\orcidlink{0000-0001-9848-9905}}
\author[c, d]{Surhud More, \orcidlink{ 0000-0002-2986-2371}}

\affiliation[a]{\CU}
\affiliation[b]{\ICRR}
\affiliation[c]{\IUCAA}
\affiliation[d]{\IPMU}

\emailAdd{gautam.bhuyan2825@gmail.com}
\emailAdd{tatha@icrr.u-tokyo.ac.jp}
\emailAdd{surhud@iucaa.in}

\abstract{The amplitude and waveform shape of gravitational wave radiation from compact binary mergers enable a direct measurement of the luminosity distance to these events, making them \textit{``standard sirens''}.
Binary black hole (BBH) mergers form a significant fraction of these standard sirens. A substantial catalog of  BBH merger events can be used to probe cosmology and large-scale structure properties, such as the baryon acoustic oscillations (BAO).
In this work, we explore detecting the BAO scale using the cross-correlation between BBH mergers and a galaxy catalog, and compare it with the BBH autocorrelation for the third-generation detector observations.
Over a $10$-year observation period, these observatories will localize approximately $1.5\times10^{5}$ events within $10$ square degrees, with \snr $\geq 50$ and luminosity distance uncertainty $\leq 50$ Mpc, and will allow recovery of the BAO scale.
We find that the BAO scale is recovered with approximately two times the \snr and a significantly better precision through cross-correlation as compared to autocorrelation.}

\begin{document}
\maketitle
\flushbottom

\section{Introduction}
Gravitational waves (GW) are ripples in the space-time of the Universe, most commonly produced by mergers of compact binary systems such as binary black holes (BBH), binary neutron stars (BNS), and neutron star-black hole binaries (NSBH). 
They are also known as \textit{``standard sirens''} because of the correlation between the detector-frame chirp mass $(\mathcal{M})$, waveform strain $(h_{\rm gw})$, and luminosity distance $(D_{L})$ of the GW event (see e.g., Ref.~\cite{Jin:2026scpma}). 
The discovery of the GW signal GW150914 in 2015 was a landmark event by the LIGO-Virgo collaboration, originating from a BBH merger event, marking the advent of the GW astronomy era~\citep{Abbott:2016prl}. 
It opened the window to independently probe the large-scale extragalactic distances and expansion history of the Universe. 
It also confirmed the existence of GWs as predicted by Einstein's theory of general relativity approximately 100 years earlier~\citep{Einstein:1916adp}. 
The detection of the BNS merger event GW170817, followed up by its electromagnetic (EM) counterparts across various instruments, kick-started the era of multi-messenger astronomy. 
Independent determination of redshift and luminosity distances from the EM and GW detectors, respectively, led to the first and so far only multi-messenger inference of the Hubble constant,
$H_{0}=71.7^{+9.4}_{-7.5}$~\kmsMpc~\citep{LIGOScientific:2026uyd}.
However, most events detected by the LIGO-Virgo-KAGRA (LVK)~\citep{Aasi:2015arxv, Acernese:2015cqgr, Akutsu:2021ptep} collaboration are primarily binary black hole (BBH) merger events~\citep{LIGOScientific:2026sit, LIGOScientific:2026wfs} which lack EM counterparts. These events are therefore referred to as \textit{dark sirens}. 
Due to the large localization uncertainty of GW events detected by current-generation detectors, the unique identification of the host galaxy is not possible~\citep{Chen:2024gdn}.
Rather, we consider the galaxies within the sky-localization area of each GW event as potential hosts to infer $H_{0}$ while marginalizing over the population model uncertainties~\citep{Mastrogiovanni:2023emh, Gray:2023wgj, LIGOScientific:2026uyd}.
This dark siren method relies strongly on the assumption of the source population model and does not utilize the spatial clustering properties of galaxies explicitly.
Another approach explores the expected clustering between GW sources and galaxies through cross-correlation to infer the redshift information~\citep{Oguri:2016dgk, Bera:2020jhx, Mukherjee:2019wcg, Ghosh:2023ksl, Ghosh:2025qwc}. 
This method is independent of the assumptions of the intrinsic population parameters that describe the GW sources.  

The current generation (2G) of ground-based GW detectors operated by the LVK collaboration are expected to improve their sensitivity as they undergo detector upgrades.
Furthermore, the upcoming third-generation (3G) detectors will provide an order-of-magnitude improvement in strain sensitivity over the 2G detectors while also allowing for probing the low-frequency GW events~\citep{Reitze:2019, Evans:2021arxv, Punturo:2010cqgr}.
These detectors are also expected to localize an order of $O(10^5)$ GW events per year within a few square degrees of sky area.
It is likely that sources of most of these events would be BBHs. 
The resulting large number of well-localized sources will therefore allow us to probe the large-scale distribution of matter in the Universe based on their spatial distribution. 
The GW source population traces the same underlying large-scale structure as the galaxies in the Universe.
Therefore, a catalog of a large number of these mergers will allow us to probe the cosmology and large-scale structure properties, such as the baryon acoustic oscillations (BAO).
It can be achieved through the two-point autocorrelations among the GW events as well as cross-correlations calculated between a catalog of GW events and a galaxy catalog.  

The Baryon Acoustic Feature (BAF) is a direct consequence of the propagation of pressure-density fluctuations in the early Universe. Prior to recombination, the temperature of the Universe was too high for stable neutral atoms to form. 
The Universe was filled with a hot, dense baryon-photon plasma. 
The photons and baryons (protons and electrons) were tightly coupled through Compton and Thomson scattering~\citep{Eisenstein:1998apj_a, Bassett:2010}. 
Although the primordial density fluctuations were initially nearly Gaussian, they grew with time and provided the seeds for the formation of large-scale structure. 
The competition between the gravitational attraction of the baryons to the dark matter overdensities and the radiation pressure of the coupled photon-baryon fluid caused these perturbations to propagate as acoustic waves through the early Universe.

As the Universe expanded and cooled, electrons and protons combined to form neutral atoms during recombination, causing the photons to decouple and the Universe to become transparent to radiation. 
The distance traveled by the acoustic waves up to this epoch, known as the sound horizon at recombination, left a characteristic scale imprinted in the spatial distribution of matter. 
After recombination, photons propagated freely to form the cosmic microwave background (CMB) observed today, while the matter perturbations continued to evolve under gravity and contributed to the formation of large-scale structure~\citep{Eisenstein:1998apj_b}. 
Consequently, the acoustic feature is imprinted in the late-time matter distribution as a characteristic comoving scale of $\sim 100,{\rm h^{-1}}$ Mpc.
This feature can be measured statistically in the clustering of galaxies and used as a ``\textit{standard ruler}'' to probe the expansion history of the Universe~\citep{Eisenstein:1998apj_b, Seo:2003apj, Blake:2003apj, DESI_DR2_II:2025prd}.

In this work, we explore the detectability of the BAF in the large-scale structure (LSS) of the Universe through GW autocorrelation and cross-correlation with a galaxy catalog. 
The paper is structured as follows: Section \ref{sec:method} describes the methodology followed in this study; Sections \ref{ssec:galaxy_catalog} and \ref{ssec:gw_catalog} deal with the simulations of the galaxy catalog and BBH merger GW events, respectively. 
Section \ref{ssec:fisher} discusses the results of the Fisher matrix estimation of the parameters of the GW events. Results of the analysis of autocorrelations and cross-correlations among the GW events themselves and the galaxy catalog, respectively, are discussed in Section \ref{ssec:corr_res}. 
Summary of the results and conclusions of this study are presented in Section \ref{sec:sumcon}.

\section{Methodology} \label{sec:method}
The density perturbations in the matter distribution field can be mathematically described using the overdensity in the matter distribution, which is defined as:
\begin{align}
    \delta (\vec{x}) & = \frac{\rho({\vec{x}})}{\overline{\rho}}-1, \label{eq:1}
\end{align}
where $\rho(\vec{x})$ denotes the matter density at the location $\vec{x}$ and $\overline{\rho}$, the comoving background matter density of the Universe, respectively (see e.g., Ref.~\citep{Cooray:2002}). 
The two-point correlation function of matter (2PCF) expresses the excess probability over random for pairs of matter particles to be separated by a distance $r$, and is given by:
\begin{align}
    \xi(\vec{r}) & = \langle \delta (\vec{x}) \delta (\vec{x}+\vec{r}) \rangle. \label{eq:2}
\end{align}
The 2PCF, $\xi(\vec{r})$, is related to the matter power spectrum of the Universe, $P(k)$, via a Fourier transform,
\begin{align}
    P(\vec{k}) & = \int d^{3}\vec{r} ~\xi(\vec{r}) e^{-i\vec{k}\cdot \vec{r}} , \label{eq:3}
\end{align}
where $\vec{k} = \frac{2\pi}{\vec{r}}$, denotes the wavenumber in units of
$h \rm{Mpc}^{-1}$  and $h = H_{0}/(100 {\rm kms}^{-1}{\rm Mpc}^{-1})$; with
$H_{0}$ denoting the Hubble constant. As we will measure the correlation
function $\xi(\vec{r})$ in narrow redshift bins, we will ignore the evolution of
the correlation function within the bin. 

In the standard cosmological model, dark matter constitutes the
dominant component of the matter density and undergoes gravitational collapse to
form halos around the peaks of the underlying matter density field. Galaxies
subsequently form within these dark matter halos, and the clustering of galaxies
therefore traces that of the underlying matter distribution, with a
scale-independent enhancement or suppression described, on sufficiently large
scales, by the linear bias factor $b_h$. The halo overdensity $\delta_h$ is
related to the total matter overdensity $\delta_{\rm m}$ through
\begin{align}
    \delta_{h} (\vec{x}, M) = \frac{n_{h}(\vec{x}, M)}{\overline{n}_{h}} -1 = b_{h}(M)\delta_{\rm m}(\vec{x}). \label{eq:4}
\end{align}
Here, $n_{h}(\vec{x}, M)$ specifies the number density of dark matter halos at
the position $\vec{x}$ having mass $M$, and $\overline{n}_{h}$ is the average number density of halos in the Universe~\citep{Cooray:2002}. As more luminous galaxies are expected to be hosted within more massive dark matter halos, they reflect the large-scale bias characteristic of their hosts, such that
\begin{align}
    b_{g} & = \int P_{h}(M) b_{h}(M) dM, \label{eq:5}
\end{align}
where $P_{h}(M)$ denotes the probability distribution of the halo masses
inhabited by the galaxies. The galaxies and gravitational wave (GW) events are
both expected to trace the underlying distribution of matter in the
Universe. However, The GW events may have their own large-scale bias $(b_{\rm
gw})$ with respect to the background matter distribution~\citep{Bera:2020apj,
Ghosh:2023ksl}. As the bias on large scales can be treated as a constant, the
cross-correlation between GW and galaxy catalogs can be used to study the
clustering properties of the large-scale structure and infer the BAF scale.

The spatial clustering between galaxies and the GW events separated by a
comoving distance $\vec{r}$, can be represented in terms of the
three-dimensional cross-correlation function $\xi_{\rm gw,g}$ following Equation
\eqref{eq:2}, such that,
\begin{align}
    \xi_{\rm gw,g} & = \langle \delta_{\rm gw}(\vec{x}) \delta_{\rm g}(\vec{x}+\vec{r})\rangle. \label{eq:6}
\end{align}
Here, $\delta_{\rm gw}$ and $\delta_{\rm g}$ represent the overdensity of the GW
sources and galaxies, respectively. The overdensity can also be expressed in
terms of the volume number density of the galaxy-GW pairs $n_{\rm gw, g}^{\rm
vol}(r)$, which can be written as:
\begin{align}
    n_{\rm gw, g}^{\rm vol}(r) & = \overline{n}_{\rm gw}^{\rm vol}
    \overline{n}_{\rm g}^{\rm vol} [1+\xi_{\rm gw,g}(r)] 4\pi r^{2}dr.
\end{align}
Here $\overline{n}_{\rm g}$ and $\overline{n}_{\rm gw}$ denote the average volume number densities of galaxy and GW source populations, respectively~\citep{Bera:2020apj}. In practice, the GW-galaxy correlation function $(\xi_{\rm
gw,g})$ can be numerically calculated based on the number of cross-correlated pair counts using the Landy-Szalay~\citep[][hereafter LS]{Landy:1993apj} estimator,
given by
\begin{align}
    \xi & = \frac{D_1 D_2 - f_{1}R_1 D_2 - f_{2}D_1 R_2 + f_{1}f_{2}R_1 R_2}{f_{1}f_{2}R_1 R_2}. \label{eq:xi_estimator}
\end{align}
Here, $D_1 D_2$ denotes the number of galaxy-GW pairs separated by a given
radius r, $R_{1}D_{2}$ represents the number of pairs between galaxies and a
random distribution of GW sources, $D_{1}R_{2}$ represents the number of pairs
between GW sources and a random distribution of galaxies, while $R_1R_2$ denotes
the number of pairs given the cross-correlation between the random backgrounds
for both galaxies and GW sources. The factors $f_{1}$ and $f_{2}$ are the ratios
of the number of objects from data and random, $f_{1(2)} =
\frac{N_{D_{1(2)}}}{N_{R_{1(2)}}}$. The same statistical framework involving the
LS estimator can also be used to estimate the GW-GW autocorrelation function,
$\xi_{\rm gw, gw}$. The BAO scale can be probed through the statistical
frameworks of GW-galaxy cross-correlation and GW-GW autocorrelation. Comparisons
between these two frameworks will be presented in Section \ref{sec:result}.

\section{Simulations} \label{sec:sim}
We investigate the potential to identify the BAO peak using two approaches: (i) the autocorrelation of GW events and (ii) the cross-correlation between GW events and galaxies.
This comparison is performed using simulated galaxy and gravitational-wave catalogs, whose constructions are described in Sections~\ref{ssec:galaxy_catalog} and~\ref{ssec:gw_catalog}, respectively.

\subsection{Galaxy Catalog Construction} \label{ssec:galaxy_catalog}

The galaxy catalog is generated using the publicly available code \lgnrm \footnote{\url{https://bitbucket.org/komatsu5147/lognormal_galaxies/src/master/}}~\citep{Agrawal:2017jcap}, which produces mock galaxy catalogs in redshift space assuming a log-normal probability density function (PDF) of galaxy and matter density fields.
We simulate a galaxy catalog with a box size of $\sim 2 {\rm ~h^{-1}Gpc}$, at the redshift $z = 0$ snapshot, adopting a linear galaxy bias of $b_{g}=1.455$ from Ref.~\cite{Agrawal:2017jcap}, and populating it with $\sim 10^{8}$ fiducial number of galaxies. 
The cosmology of the Universe is adopted from Ref.~\cite{Planck:2016anda}.
The input power spectrum used for the galaxy catalog simulation is computed using the Eisenstein-Hu transfer function~\citep{Eisenstein:1998apj_a} and includes the BAO feature at $100 {\rm h^{-1}}$Mpc scale.
The \lgnrm catalog provides the three dimensional comoving coordinates $(x, y,z)$  in the units of $\rm{h}^{-1}$Mpc scale along with the corresponding velocity components $(v_{x}, v_{y}, v_{z})$ in km~s$^{-1}$.
Assuming the observer is located at the center of the simulated catalog with a box size of $2 h^{-1}{\rm Gpc}$, we select galaxies within a spherical region of radius $1 {\rm ~h^{-1}Gpc}$. 
However, the \lgnrm catalog does not provide any information about the brightness of these galaxies. 
Galaxy brightness or magnitudes (either observed or absolute) are crucial in selecting the galaxies observable at a given redshift. 
Galaxies fainter than an absolute magnitude limit may not be observable or detectable to the observer. 
Therefore, we need to construct a flux-limited or absolute magnitude-limited galaxy catalog containing only the observable galaxies at a given maximum redshift defined by the size of the galaxy catalog.
We use the galaxy magnitudes available from the \mice catalog~\footnote{\mice is a publicly available mock galaxy survey catalog generated from the halo catalogs available from the Marenostrum Institut de Ci\`encies de l’Espai (MICE) simulation~\citep{Carretero:2017pepschep}.} to construct a flux-limited catalog from the \lgnrm catalog.  
We consider a subset of $1/8^{\rm th}$ of the total \mice catalog galaxy sample size~\citep{Tallada:2020ac, Crocce:2015mnras, Carretero:2017pepschep}. 
It utilizes the halo occupation distribution (HOD) and sub-halo abundance matching (SHAM) algorithm to generate the galaxy catalog within the redshift range: $0.07296  < z < 1.4$. 
It covers a total sky area of $\sim 5000 ~{\rm deg^{2}}$, or one octant of the total sky, without repeating the simulation box. 
{\small \mice} is complete for DES-like surveys $(m_{\rm i} < 24)$ out to the redshift $z<1.4$~\citep{Crocce:2015mnras}.  

\begin{figure*}
\centering
    \includegraphics[width=1\linewidth, keepaspectratio]{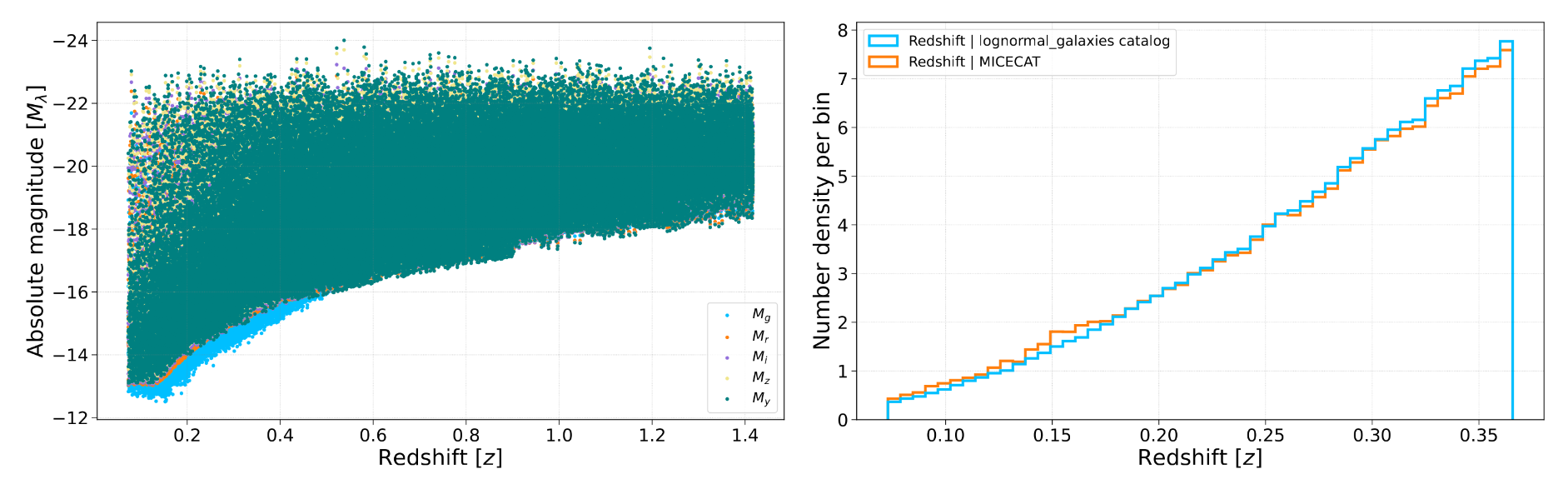}
    \caption{The \textit{left panel} shows the distribution of absolute magnitudes of the galaxies in {\scriptsize \textsc{micecat}} catalog as a function of their redshifts. The \textit{right panel} shows the distribution of redshifts of the galaxies sampled from the simulated \lgnrm catalog to match the redshift distribution of galaxies in the {\scriptsize \textsc{micecat}} catalog.}
    \label{fig:flux_limit}
\end{figure*}

We follow the following steps to construct a flux-limited galaxy catalog using \mice galaxy magnitudes:
\begin{enumerate}

    \item We determine the redshift distribution of galaxies inside $1\rm{h}^{-1}$Gpc sphere of the \lgnrm catalog. We estimate galaxy redshifts from their comoving distances assuming the cosmology~\cite{Planck:2016anda}. Galaxies corresponding to the sphere of size $1\rm{h^{-1}}$Gpc have redshifts in the range: $0 \leq z < 0.4$. We therefore consider \mice galaxies with redshifts $z \lesssim 0.4 $.

    \item \textit{Left panel} in Figure \ref{fig:flux_limit} shows the distribution of absolute magnitudes of the galaxies in the \mice catalog as a function of their redshifts.
    It is observed from the figure that the galaxies fainter than -16 mag in the DES $g-$band in the \mice catalog may not be detectable at redshifts $z \sim 0.4$.
    Therefore, we only consider the \mice galaxies brighter than -16 mag in the DES $g-$band in our analysis in the redshift range $0 < z \lesssim 0.4$.  
    
    \item Galaxies within the $1\rm{h}^{-1}$Gpc spherical volume of the \lgnrm catalog are then sampled based on their estimated redshift to match the redshift distribution of galaxies in the flux-limited \mice catalog. We use the rejection sampling algorithm for this purpose. 
    The resulting redshift distribution is shown in Figure \ref{fig:flux_limit}, along with the redshift distribution of the \textsc{mice} catalog, demonstrating good agreement between them.
\end{enumerate}
The final \lgnrm catalog galaxies span the redshift range $0.07296 \leq z < 0.4$, with a distribution consistent with that of the \mice galaxy redshifts. This galaxy catalog is then used to compute the GW-galaxy cross-correlations.

\subsection{GW signals using 3G detectors} \label{ssec:gw_catalog}
The upcoming third-generation (3G) GW observatories, such as the Cosmic Explorer (CE)~\citep{Reitze:2019, Evans:2021arxv} and the Einstein Telescope (ET)~\citep{Punturo:2010cqgr}, are expected to achieve an order of magnitude improvement in sensitivity compared to current-generation detectors. 
Although the final locations of these detectors have not yet been finalized, their fiducial locations are commonly adopted for numerical simulations of GW signals and have been widely used in previous studies~\citep{Kumar:2022apj, Nitz:2021apjl}.
In this work, we consider a two-detector network configuration of one CE in the USA and ET at the same location as Virgo. The fiducial locations and design parameters of the GW detectors used in our analysis are summarized in Table~\ref{tab:detector locations}, following Ref.~\cite{Nitz:2021apjl}.

The 3G GW detectors will improve the localizations of BBH and BNS events significantly. Although the BNS events typically have higher intrinsic merger rates, the majority of GW events detected in past observations by the LVK detector configuration are the BBH events~\citep{LIGOScientific:2026sit}. However, the horizon distance for BBH is much larger than that for BNS. Therefore, improved localization due to the sensitivity of the 3G detectors will only result in a significant number of detected BBH events with a higher probability of astrophysical compact binary coalescence origin, $p_{\rm astro} >0.5$~\citep{LIGOScientific:2026sit}. 

\begin{table*}[]
    \caption{Location, noise curve, and low-frequency cutoff $f_{\rm low}$ used for the detector configurations considered in this work.}
    \centering
    \begin{tabular}{cccccc}
    \hline
    \hline
     Abbreviation & Observatory  &  $f_{\rm low}$ (Hz) & Noise curve & Latitude (deg) & Longitude (deg)\\
     \hline
      $C^{U}_{1}$ & Cosmic Explorer USA & $5.2$ & CE1 & $40.8$ & $-113.8$ \\
      $E$ & Einstein Telescope & $2$ & ET-D Design & $43.6$ & $10.5$ \\
    \hline
    \hline
    \end{tabular}
    \label{tab:detector locations}
\end{table*}

Ideally, Bayesian parameter estimation should be performed to infer the source parameters from GW signals. However, in this study, we consider next-generation detectors, for which the expected detection of a large number of events becomes computationally expensive. 
We therefore adopt the Fisher-matrix approximation to the GW likelihood as an efficient approach for estimating source-parameter uncertainties in the high-\snr limit \citep{Vallisneri:2007ev}. 
We use the Fisher matrix formalism implemented by the \textsc{gwfish} software package \citep{Dupletsa:2023grqc}, which is based on a Gaussian approximation of the likelihood function:
\begin{align}
    \mathcal{L}(\chi) \propto \exp{(- \vec{\chi}^{\rm T} \mathcal{C}^{-1} \vec{\chi}/2)}, \label{eq:8}
\end{align}
where $\vec{\chi}$ denotes the parameter set and $\mathcal{C}$ denotes the covariance matrix, respectively. The Fisher matrix $\mathcal{F}$ is the inverse of the covariance matrix, whose components can be written as:
\begin{align}
    \mathcal{F}_{ij} & = \sum_{k=1}^{N} \langle \partial_{\chi_{i}} h^{k}|\partial_{\chi_{j}} h^{k} \rangle. \label{eq:9}
\end{align}
Here, $h(\vec\chi)$ represents the waveform model, and $N$ is the number of components of the GW detector network. The derivatives of the waveform are calculated with respect to the model parameters such as chirp mass $(\mathcal{M})$ and mass ratio $(q \leq 1)$ of the BBH systems, luminosity distance $(D_{\rm L})$, RA, Dec, etc., represented as $\vec{\chi} = (\chi_{i})$. The inner product is defined on the signal-model manifold as :
\begin{align}
    \langle a|b\rangle & = 4\int_{0}^{\infty} df \frac{\Re\!\left(a(\vec{\chi},f) b^{*}(\vec{\chi},f)\right)}{S_{n}(f)}, \label{eq:10}
\end{align}
where $S_{n}(f)$ denotes the instrument noise spectral density required by the model.

We randomly sample $\sim 3 \times 10^{5}$ galaxies as hosts of BBH mergers from the galaxy catalog before applying any flux or redshift limit. The redshifts and sky positions (right ascension and declination) of these galaxies are used to assign the locations of the GW events. 
The BBH component source-frame masses $(m_{1}, m_{2})$, with mass ratio $q=m_{2}/m_{1} \leq 1$, are sampled from \textsc{Power Law+Gaussian} distribution between minimum mass $5~M_{\odot}$ and maximum mass $60~M_{\odot}$.
The other BBH parameters are randomly sampled from the default parameter distributions of \textsc{bilby} (see Table 1 of Ref.~\cite{Ashton:2019apjs}).
We consider a uniform distribution for the merger time $t_{c}$ over the 10-year observation period.
We then simulate the corresponding BBH GW signals and consider events with a network \snr $\geq 10$ to be detected.
We employ the Fisher matrix formalism to estimate the uncertainties in the BBH parameters.
The Fisher-matrix analysis includes the chirp mass $(\mathcal{M})$, mass ratio $(q)$, luminosity distance $(D_{\rm L})$, right ascension (RA), declination (Dec), and the angle between the line of sight to the observer and the total angular momentum of the binary system $(\theta_{JN})$. The covariance matrix of these parameters is obtained from the inverse of the Fisher matrix.

Based on the covariance matrix of the BBH source parameters, we construct their multivariate normal distributions and draw $100$ random samples of the parameters: $D_{L}$, RA, and Dec of each BBH event. 
We randomly select 10 such samples from these randomly drawn parameter samples of each BBH event.
This procedure is repeated for $1000$ different realizations of the injected BBH events in the galaxy catalog, to finally obtain $10^{4}$ such sub-realizations of the BBH population.
The autocorrelation among the GW events and cross-correlation with the galaxies are computed for each such sub-realization.
The $1\sigma$ uncertainties of the correlation functions are estimated from the $16^{\rm th}$, $50^{\rm th}$, and $84^{\rm th}$ percentiles of the values obtained for all the sub-realizations.

\section{Results}\label{sec:result}

We begin by describing the error estimates of the GW signal parameters using the Fisher matrix formalism in Section \ref{ssec:fisher}. We then assess the prospects for identifying the BAO peak from the autocorrelation of GW events versus the cross-correlation between galaxies and GW events (Section~\ref{ssec:corr_res}).

\subsection{Fisher-Matrix Estimates of Gravitational-Wave Parameters}\label{ssec:fisher}

\begin{figure}[hbt!]
    \centering
    \includegraphics[width=\linewidth, keepaspectratio]{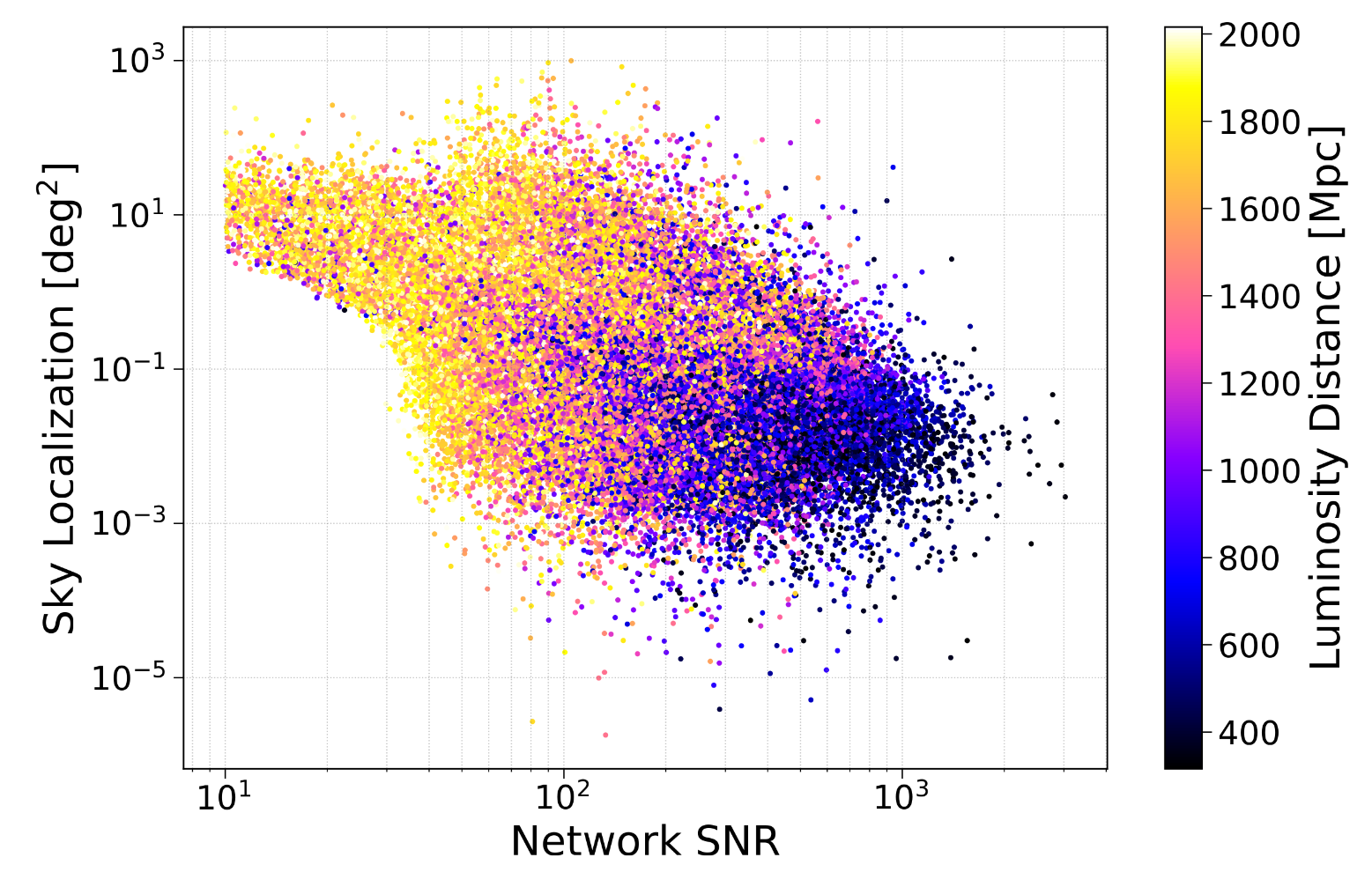}
    \caption{The $90^{\rm th}$ percentile errors in sky localization area of the simulated GW events as a function of the signal-to-noise ratio (SNR) of the detector network, color-coded with the luminosity distance of the simulated GW events.}
    \label{fig:2}
\end{figure}

In this section, we briefly describe the measurement uncertainties associated with the GW signal parameters. We have estimated the Fisher errors for RA, Dec, $D_{L}$, $\mathcal{M}$, $q$, and $\theta_{JN}$ for $\sim 3\times 10^{5}$ events. Figure \ref{fig:2} presents the $90^{\rm th}$-percentile sky-localization area as a function of the detector-network \snr, with each event color-coded according to its luminosity distance. The figure shows that most of the events with high network \snr are localized within $10 ~{\rm deg^{2}}$. 
We therefore apply the following selection criteria: network \snr $\geq 50$, $90^{\rm th}$ percentile sky-localization area $\leq 10 ~{\rm deg^{2}}$, and luminosity-distance uncertainty $\sigma_{D_{\rm L}} \leq 50 $ Mpc. Approximately $1.5\times 10^{5}$ out of the injected $\sim 3\times 10^{5}$ events satisfy these selection criteria. 
We construct catalogs of GW events by drawing posterior samples from the distributions of $D_{L}$, RA, and Dec based on the respective $1\sigma$ error bars of each qualifying event obtained from the computation of the Fisher matrix parameters. 

\subsection{Correlation analysis}\label{ssec:corr_res}
\begin{figure}[hbt!]
    \centering
    \includegraphics[width=\linewidth, keepaspectratio]{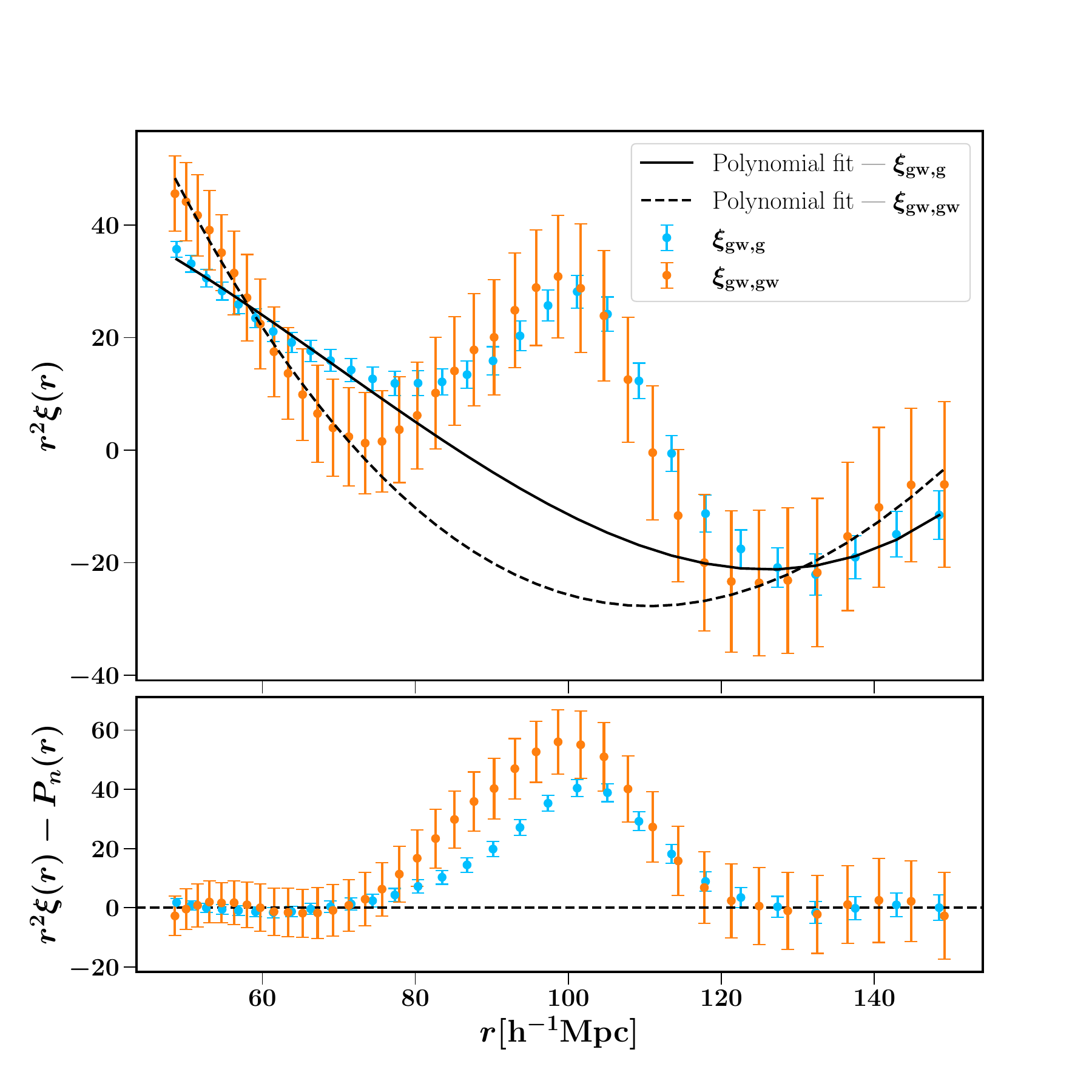}
    \caption{\textit{Top panel} shows the comparison of the recovered BAO peak and the error bars obtained using autocorrelation between the GW events and cross-correlation between the galaxies and GW events. The black colored continuous and dashed lines represent the polynomials fitted to the cross- and autocorrelation functions, respectively, to determine the \snr of the BAO signals obtained using these two methods. \textit{Bottom panel} shows the residuals of between the fitted function $P_{n}(r)$ and the scaled correlation functions $r^{2}\xi$.}
    \label{fig:3}
\end{figure}

We compute the two-point autocorrelation of GW events using Eq.~\eqref{eq:xi_estimator} by counting pairs in $30$ concentric shells in comoving separation with radii $r \in [60, 140]$ around each event.
We also compute the cross-correlation between galaxies and GW events by counting pairs around each GW event within the same set of shells. 
The pair counts for constructing the correlation functions have been carried out using the \texttt{CKDTree.count\_neighbors} method available from \texttt{scipy.spatial.CKDTree} \textsc{Python} package.  
Figure \ref{fig:3} compares the recovered BAO peak and associated $1\sigma$ error bars obtained from the autocorrelation of GW events and cross-correlation between the galaxies and GW events. 
We observe that the $1\sigma$ uncertainties on the BAO peak recovered from the autocorrelation are significantly larger than those using the cross-correlation in each comoving distance bin. 
We also calculate the \snr of the recovered BAO peak from both methods discussed here by fitting a third-order polynomial to the region of auto- and cross-correlation estimates, which do not contain the BAO peak, using:
\begin{align}  
    \mathrm{\snr} & = \sqrt{\left[r^{2}\xi - P_{n}(r)\right]^{T} C^{-1} \left[r^{2}\xi - P_{n}(r)\right]}. \label{eq:snr}
\end{align}
Here, $P_{n}(r)$ and $C$ represent the fitted polynomial function and covariance matrix constructed with the squared $1\sigma$ error bars along the diagonal.
To determine the location of the BAO peak in the comoving space, we fit the correlation functions with a polynomial added to a Gaussian, which can be written as:
\begin{align}
    f(r) & = P_{n}(r) + A\exp\left[-\frac{(r-r_{d})^{2}}{2\sigma_{d}^{2}}\right], \label{eq:fit}
\end{align}
where $r_{d}$ and $\sigma_{d}$ represent the location of the BAO peak and its width, respectively. 
We follow a two-step process to fit the BAO peak in the correlation function.
As before, the polynomial $P_{n}(r)$ is first fitted to the regions of the correlation function that exclude the peak. 
The weighted least-squares minimization technique has been used to fit the polynomial function, with the weights given by the inverse variance-covariance matrix of the correlation function.
In the second step, we fit the residuals of the polynomial fit with a Gaussian using the Bayesian Markov Chain Monte Carlo (MCMC) technique to determine the location of the BAF.
We find that the BAO peak is recovered at a significantly higher \snr by the cross-correlation, ${\rm \snr_{\rm gw, g} } \approx 1.58 \times 10^{5}$ between galaxies and the GW events, as compared to the autocorrelation among the GW events, ${\rm \snr_{\rm gw, gw} } \approx 8.49\times 10^{4}$.

\begin{figure*}[hbt!]
    \centering
    \begin{tabular}{cc}
    \resizebox{0.5\linewidth}{!}{
    \includegraphics[width=0.5\linewidth, keepaspectratio]{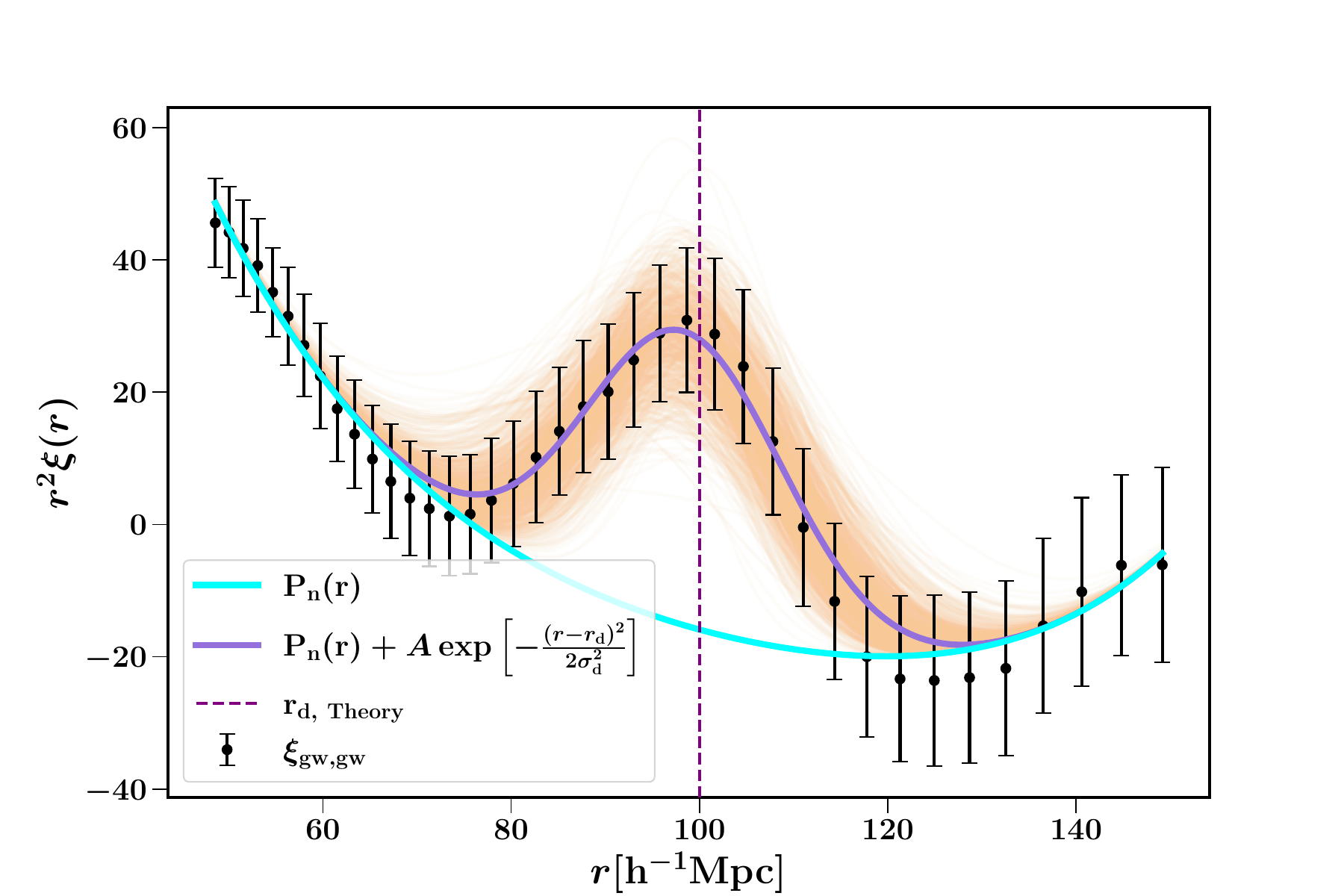}
    }  &  
    \resizebox{0.5\linewidth}{!}{
    \includegraphics[width=0.5\linewidth, keepaspectratio]{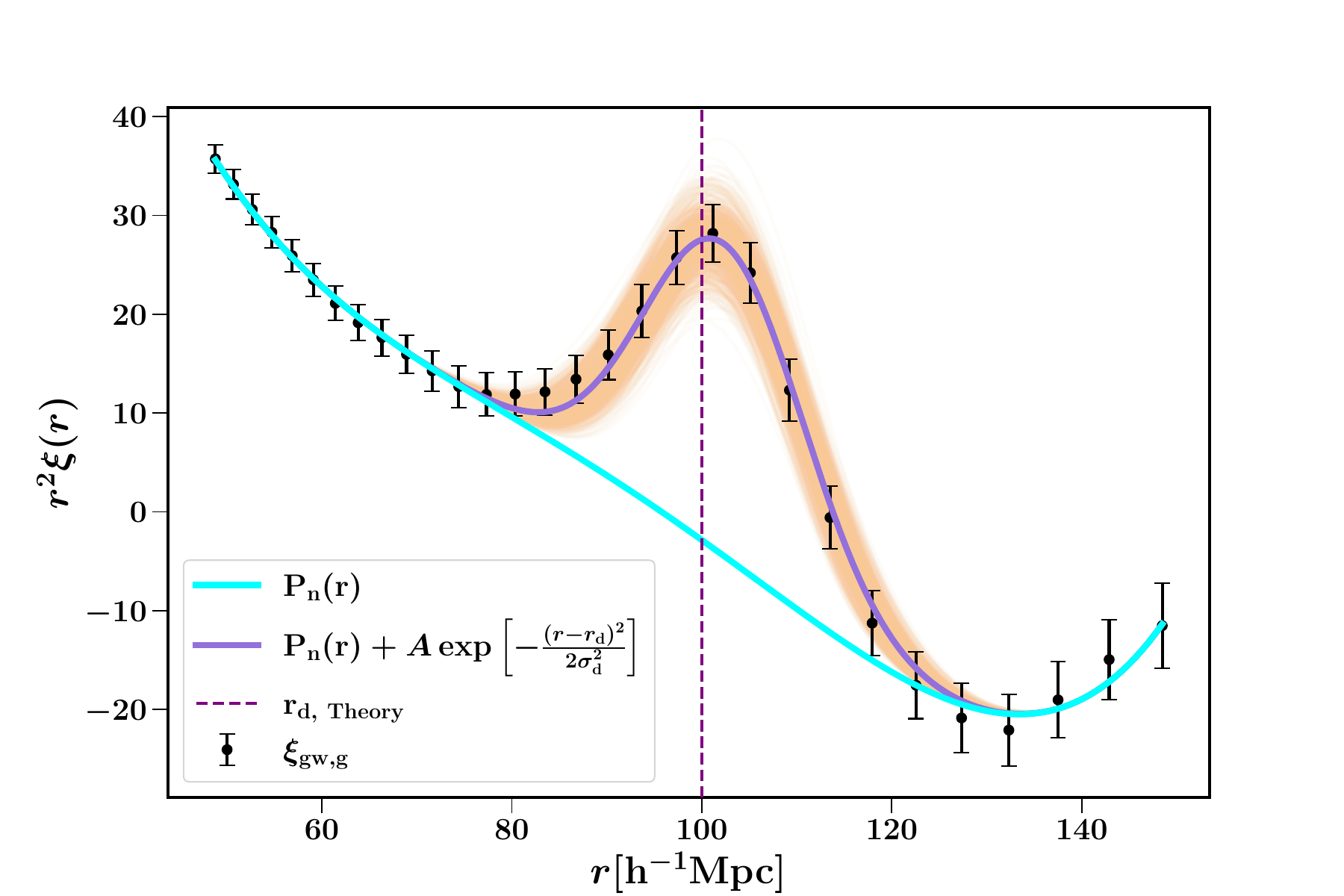}
    }
    \end{tabular}
    \caption {The BAO peak recovered by the GW auto-correlation (\textit{left panel}) and cross-correlation (\textit{right panel}) with galaxies. 
    They are fitted with a polynomial added to a Gaussian to determine the location of the BAO peak.
    The cyan colored curves on both panels represent the polynomial $P_{n}(r)$ fitted to the correlation functions.
    The purple curves represent the fitted \textsc{Gaussian + Polynomial} full model using the MCMC best-fit parameters.
    The vertical dashed line on both panels denotes the injected value of $r_{d}$.}
    \label{fig:4}
\end{figure*}

Figure \ref{fig:4} shows the BAO peak recovered by the cross-correlation between the galaxies and GW events fitted with the function $f(r)$ represented by Equation \eqref{eq:fit}. The location of the BAO peak is found to be $r_{d} \approx 98.33^{+ 2.49}_{-2.42} {~\rm h}^{-1}$Mpc and  $102.34^{+0.84}_{-0.84} {~\rm h}^{-1}$Mpc, respectively through auto-correlation and cross-correlation. We find that the BAO peak is localized with greater precision by cross-correlation between GW and galaxy catalogs as compared to autocorrelation among the GW events. We present the results of the polynomial fit and polynomial + Gaussian fit in Table \ref{tab:2}. 

\begin{table}[hbt!]
    \caption{\snr and the location of the BAO peaks recovered by GW autocorrelation and cross-correlations with galaxies.}
    \centering
    \begin{tabular}{cccc}
    \hline
    \hline
    Method & \snr & $r_{d} \pm \sigma_{r_d}$ & BAO peak width $(\sigma_{d})$  \\
    &&$[\mathrm{h}^{-1}$Mpc]&$[\mathrm{h}^{-1}$Mpc] \\
    \hline
    $\xi_{\rm gw,gw}$ & $8.49\times 10^{4}$ & $98.33^{+ 2.49}_{-2.42}$ & $10.53^{+2.02}_{-2.86}$ \\
    $\xi_{\rm gw, g}$ & $1.58 \times 10^{5}$ & $102.34^{+0.84}_{-0.84}$ & $8.38^{+0.74}_{-0.83}$ \\
    \hline
    \hline
    \end{tabular}
    
    \label{tab:2}
\end{table}

\section{Summary and Conclusions} \label{sec:sumcon}
In this work, we explore the detectability of the BAO scale using autocorrelation 
among the GW wave events from BBH mergers and the cross-correlation of these events with 
galaxies. We use a 3G detector network of ET and a CE each to simulate the BBH merger 
events using the Fisher matrix formulation, which is computationally less expensive than 
full parameter estimation for these kinds of studies. The low number density of detected 
GW events and the large uncertainty associated with their sky localization make it very 
challenging to probe the properties of the large-scale structure, such as the BAO scale. 
However, with the 3G detector network configuration, we find that there are 
approximately $1.5\times10^{5}$ events localized within $10$ square degrees of sky-
localization area, with high \snr $(\geq 50)$, and lower luminosity distance uncertainty 
$(\leq 50$ Mpc), in 10 years of observation period. A catalog of these events is used to 
calculate the autocorrelation and cross-correlation functions, using the Landy-Szalay 
\citep*{Landy:1993apj} estimator. The results obtained in this study are as follows:

\begin{enumerate}
    \item The BAO peak is detected through both autocorrelation among the GW events and cross-correlation with the galaxies.
    \item The BAO scale is recovered by the cross-correlation with galaxies with significantly high \snr $(\sim 1.58 \times 10^{5})$ as compared to the autocorrelation (\snr $\sim 8.48 \times 10^{4}$). 
    \item A polynomial of order $n=3$ added to a Gaussian has been fitted to the correlation functions to determine the size of the BAO scale. 
    \item We find that the BAO scale is recovered with greater precision through cross-correlation of GW catalogs with a galaxy catalog ($r_{d} \approx 102.34^{+0.84}_{-0.84} {~\rm h}^{-1}$Mpc) as compared to autocorrelation among the GW events ($r_{d} \approx 98.33^{+ 2.49}_{-2.42} {~\rm h}^{-1}$Mpc). 
\end{enumerate}

Our results show that a large number of high \snr BBH events could be detected within 
$10$ sq. degree sky localization area and small $D_{L}$ errors with a 3G two-detector 
network configuration. We effectively quantify that the cross-correlation between a 
catalog of GW events and a galaxy catalog serves as a tool to probe the BAO scale with 
greater precision. It will lead to the determination of the value of $H_{0}$ and other 
cosmological parameters with greater precision in an independent manner. 
The upcoming galaxy surveys with increasing completeness at higher redshifts will allow 
us to probe the properties of the large-scale structure through the GW-galaxy cross-
correlation method. 

\acknowledgments
GB is grateful to his PhD supervisor, Dr. Sukanta Deb, and IUCAA for supporting his
academic visits to IUCAA for carrying out this work through the IUCAA visiting associate 
program. 
GB is grateful to the Department of Science and Technology (DST), Govt. of India,
New Delhi, for providing the financial support for this study as a Senior Research
Fellow (SRF) through the DST INSPIRE Fellowship research grant 
(DST/INSPIRE/Fellowship/2019/IF190616).
T.G. acknowledges the support from the JSPS Grant-in-Aid for Transformative Research Areas (A) No. 23H04893.
We acknowledge the use of the IUCAA LDG cluster Sarathi and PEGASUS cluster for the
computational/numerical work. 

\bibliographystyle{JHEP}
\bibliography{sample631}

\end{document}